\documentclass[preprint,12pt, a4paper]{elsarticle}

\usepackage{amssymb}
\usepackage{amsmath}

\usepackage{lineno}

\usepackage{import}
\usepackage{xfrac}
\usepackage{pgf}
\usepackage{float}
\usepackage{listings}
\usepackage[breaklinks=true]{hyperref}
\usepackage{breakurl}
\usepackage[utf8x]{inputenc}

\lstdefinestyle{BashInputStyle}{
	language=bash,
	basicstyle=\small\sffamily,
	numbers=left,
	numberstyle=\tiny,
	numbersep=3pt,
	frame=tb,
	columns=fullflexible,
	backgroundcolor=\color{yellow!20},
	linewidth=0.9\linewidth,
	xleftmargin=0.1\linewidth
}

\newcommand*{\Package}[1]{\texttt{#1}}%

\journal{SoftwareX}

\begin{document}

\begin{frontmatter}



\title{\makebox[0pt]{PlenoptiSign: an optical design tool for plenoptic imaging}}


\author{Christopher Hahne, Amar Aggoun}

\address{University of Wolverhampton, School of Mathematics and Computer Science, Wolverhampton, United Kingdom}
\begin{abstract}
Plenoptic imaging enables a light-field to be captured by a single monocular objective lens and an array of micro lenses attached to an image sensor. %
Metric distances of the light-field's depth planes remain unapparent prior to acquisition. Recent research showed that sampled depth locations rely on the parameters of the system's optical components. This paper presents PlenoptiSign, which implements these findings as a Python software package to help assist in an experimental or prototyping stage of a plenoptic system.

\end{abstract}


\end{frontmatter}




\section{Motivation and significance}
\label{}
Plenoptic cameras gain increasing attention from the scientific community and pave their way into experimental three-dimensional (3-D) medical imaging~\cite{prevedel:2014:simultaneous, Li439315, Bedard:17, Palmer:18}. 
A limitation of light-field cameras is that the maximum distance between two viewpoint positions, the so-called baseline, is confined to the extent of the entrance pupil~\cite{AW, Hahne:IJCV:18}.
With triangulation, small baselines are mapped to depth planes close to the imaging device making them suitable for medical purposes \if depth sensing instruments\fi such as in a microscope~\cite{prevedel:2014:simultaneous, Li439315} or an otoscope~\cite{Bedard:17}. %

When capturing depth with a plenoptic system, it is essential to place available light-field depth planes on the targets of interest. Hence, it is a key task in 
plenoptic data acquisition to choose suitable specifications for the micro lenses, the objective lens and the sensor early to save time and costs at the conceptual design stage of a prototype. 

\Package{PlenoptiSign} enables {\it a priori} depth plane localization in plenoptic cameras for stereo matching from sub-aperture image disparities~\cite{AW, NG, Hahne:IJCV:18} as well as computational refocusing via shift-and-integration~\cite{Isaksen:2000:DRL:344779.344929}. 
 \Package{PlenoptiSign} can be used to pinpoint object distances in light-field images rendered by our complementary open-source software \Package{PlenoptiCam}~\cite{Plenopticam:2019}.
The underlying physical model of \Package{PlenoptiSign} was devised and experimentally proven in studies by Hahne~{\it et al.}~\cite{Hahne:IJCV:18, Hahne:OPEX:16} and applies to the Lytro-type setup~\cite{LYTROMAN} at this stage. %

Given an experiment involving plenoptic image acquisition, a researcher may want to simulate the influence of the Micro Lens Array (MLA), the objective lens and its focus as well as optical zoom settings to investigate the depth resolution performance. With this software, the user is able to optimize an experimental setup as required. In its current state, the tool can be called from a Graphical User Interface~(GUI), a web-server capable of handling the Common Gateway Interface (CGI) or from the Command Line Interface~(CLI) where a user is asked for all input parameters.

It is the goal of this paper to raise awareness of the light-field model's capabilities when implemented as a software tool. 
In Section~\ref{sec:2}, we sketch the software architecture while turning the focus onto the ray function solver by means of linear algebra to complement previous publications where this has not been explained in much detail. This is followed by usage instructions and exemplary result presentations in Section~\ref{sec:3} after which the potential influence on future applications is discussed in Section~\ref{sec:4}. 

\section{Software description}
\label{sec:2}


\subsection{Architecture}
\label{sec:2:1}

\begin{figure}[H]
	\centering
	\includegraphics[scale=.37]{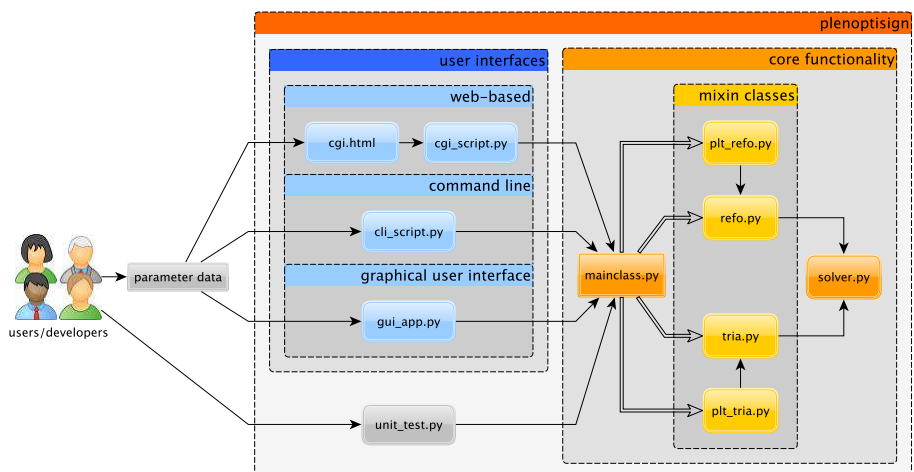}
	\caption{PlenoptiSign's software architecture\label{fig:arch}}
\end{figure}

Plenoptic camera parameters are passed to \Package{PlenoptiSign} using either the CLI ({\it cli\textunderscore script.py}), a \Package{tkinter} GUI ({\it gui\textunderscore app.py}) or a CGI web-server ({\it cgi\textunderscore script.py}). Based on Python's \Package{ddt} package, a unit test was added to support potential future development. An overview of the code structure is depicted in Fig.~\ref{fig:arch}. An object of the light-field geometry estimator is instantiated from a user interface by calling {\it mainclass.py}, which inherits mixin classes, namely {\it refo.py}, {\it tria.py}, {\it plt\textunderscore refo.py} and {\it plt\textunderscore tria.py}.

\subsection{Functionalities}
\label{subsec:2:2}

The core task of this software is to find intersections of light-field ray pairs. Since its implementation has not been entirely covered before~\cite{Hahne:IJCV:18, Hahne:OPEX:16}, it is demonstrated \if in full detail\fi hereafter. A concise form of finding an intersection of two ray functions is by solving a System of Linear Equations~(SLE) given as
\begin{align}
\mathbf{A}\mathbf{x}=\mathbf{b}
\end{align}
where $\mathbf{A}$ and $\mathbf{b}$ make up ray functions and $\mathbf{x}$ contains unknowns representing locations of ray intersections in the $y-z$ plane. Using two ray functions, we obtain a unique solution to the SLE by the algebraic inverse $\mathbf{A}^{-1}$
\begin{align}
\mathbf{x}=\mathbf{A}^{-1}\mathbf{b}
\end{align}
where $\mathbf{A} \in \mathbb{R}^{2 \times 2}$ is invertible.
To cover cases of overdetermined SLEs, a more generic solution is provided by the Moore-Penrose pseudo-inverse $\mathbf{A}^{+}$
\begin{align}
\mathbf{x}&=\mathbf{A}^{+}\mathbf{b} \\
\mathbf{A}^{+} &= (\mathbf{A}^\intercal \mathbf{A})^{-1}\mathbf{A}^\intercal
\end{align}
with $^\intercal$ denoting the matrix transpose.
Algebraic computations are implemented by means of the \Package{NumPy} module in the {\it solver.py} file.

\begin{figure}[H]
	\centering
	\includegraphics[width=\linewidth]{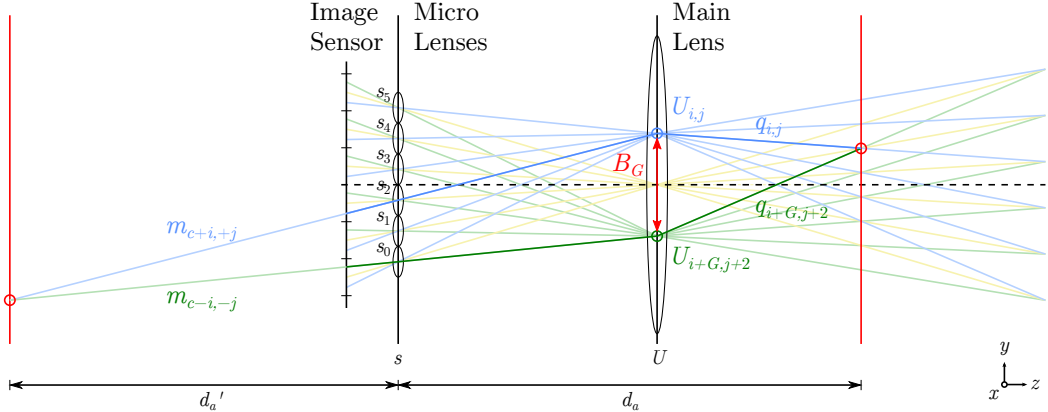}
	\caption{Plenoptic model with simplified MLA plane $s$ and main lens $U$ along which the baseline $B_G$ ($G=2$) is found via intersections $U_{i,j}$ from object-side ray slopes $q_{i,j}$. Refocused object distances $d_a$ rely on the image distance extension ${d_a}'$ obtained from image-side ray slopes $m_{i,j}$ and micro lens positions $s_j$. %
	\label{fig:model}}
\end{figure}

For plenoptic triangulation, matrices $\mathbf{A}$ and $\mathbf{b}$ may be defined according to the notation provided in~\cite{Hahne:IJCV:18} with Eq.~(18) as a ray function, which writes
\begin{align}
\widehat{f}_{i, \, j}(z) &= q_{i, \, j} \times z+U_{i, \, j} \, \, ,\quad z \in \left[U,\infty\right)
\end{align}
with $U_{i, \, j}$ as $y$-intercepts at the object-side principal plane and $q_{i, \, j}$ as chosen chief ray slopes in object space. Here, $i$ is an arbitrary micro image pixel index and $j$ represents a micro lens index in one direction. As seen in Fig.~\ref{fig:model}, we find the separation of two light-field viewpoints, so-called baseline $B_G$, by two intersecting linear ray functions~\cite{Hahne:IJCV:18}
\begin{align}
B_G := \widehat{f}_{i, \, j}(z) &= \widehat{f}_{i+G, \, j+1}(z) \label{eq:rayTracingEq2}
\end{align}
where our reference viewpoint depends on $i$ and is separated by a scalar $G$ of light-field viewpoints. After rearranging the two functions, we write
\begin{align}
-q_{i, \, j} \times z + B_G &= U_{i, \, j} \\
-q_{i+G, \, j+1} \times z + B_G &= U_{i+G, \, j+1}
\end{align}
which can be solved using the SLE given by
\begin{align}
\mathbf{A} =
\begin{bmatrix}
- q_{i, \, j} & 1 \\
- q_{i+G, \, j+1} & 1 \\
\end{bmatrix}
; \quad
\mathbf{x} = 
\begin{bmatrix}
z \\
B_G \\
\end{bmatrix}
; \quad
\mathbf{b} =
\begin{bmatrix}
U_{i, \, j} \\
U_{i+G, \, j+1} \\
\end{bmatrix}
\end{align}
where $z$ here corresponds to the longitudinal entrance pupil position $\overline{A''H_{1U}}$~\cite{Hahne:IJCV:18}. 
\newpage

Similarly, we can trace a pair of light-field rays to find an object plane to which a plenoptic image is focused using the shift and integration algorithm~\cite{Isaksen:2000:DRL:344779.344929, Hahne:OPEX:16}. In accordance with the model presented in~\cite{Hahne:OPEX:16}, an image-side ray function $f_{c+i, j}(z)$ is given by
\begin{align}
f_{c+i, j}(z) = m_{c+i, j} \times z+s_j, \quad z \in(-\infty, U] \label{eq:rayTracingEq1}
\end{align}
where $m_{c+i, j}$ denotes an image-side ray slope and $s_j$ the respective micro lens position.
A ray is chosen by %
$i=-c$ and $j=a(M-1)/2$ with its counterpart ray having negative indices. Applying this to Eq.~\ref{eq:rayTracingEq1} and rearranging yields
\begin{align}
-m_{c+i, +j} \times z = s_{+j} \\
-m_{c-i, -j} \times z = s_{-j}
\end{align}
which can be represented in matrix form by
\begin{align}
\mathbf{A} =
\begin{bmatrix}
- m_{c+i, \, +j} & 1 \\
- m_{c-i, \, -j} & 1 \\
\end{bmatrix}
; \quad
\mathbf{x} = 
\begin{bmatrix}
{d_a}' \\
y \\
\end{bmatrix}
; \quad
\mathbf{b} =
\begin{bmatrix}
s_{+j} \\
s_{-j} \\
\end{bmatrix}
\end{align}
where $y$ is the vertical position and ${d_a}'$ is an elongation of the image distance which is mapped to respective refocusing distance $d_a$ using the thin lens equation while taking thick lenses and their principal planes \if$H_{1U}$, $H_{2U}$\fi into account~\cite{Hahne:OPEX:16}.
%
%
\section{Illustrative Examples}
\label{sec:3}
\subsection{Usage}
After downloading \Package{PlenoptiSign} from the \href{https://github.com/hahnec/plenoptisign}{online repository}, installation is made possible with the \Package{setuptools} module which can be run by

\lstinline[style=BashInputStyle]'\$ python plenoptisign/setup.py install' \\
in the download directory, provided that Python is available and granted necessary privileges. As indicated in Fig.~\ref{fig:arch}, \Package{PlenoptiSign} can be accessed from three different interfaces, namely web-based CGI, a GUI and bash command line. Executing \Package{PlenoptiSign} from the command line is done by

\lstinline[style=BashInputStyle]'\$ plenoptisign -p' \\
where \lstinline[style=BashInputStyle]'-p' sets the plot option with rays and depth planes being depicted as seen in Figs.~\ref{fig:refo} and \ref{fig:3d_tria}. Starting the \Package{tkinter}-based GUI application is either done by running a bundled executable file or by

\lstinline[style=BashInputStyle]'\$ plenoptisign -g'\\
For more information on available commands, use the help option \lstinline[style=BashInputStyle]'-h'.

\subsection{Results}
\begin{figure}[H]
	\centering
	\resizebox{\textwidth}{!}{\input{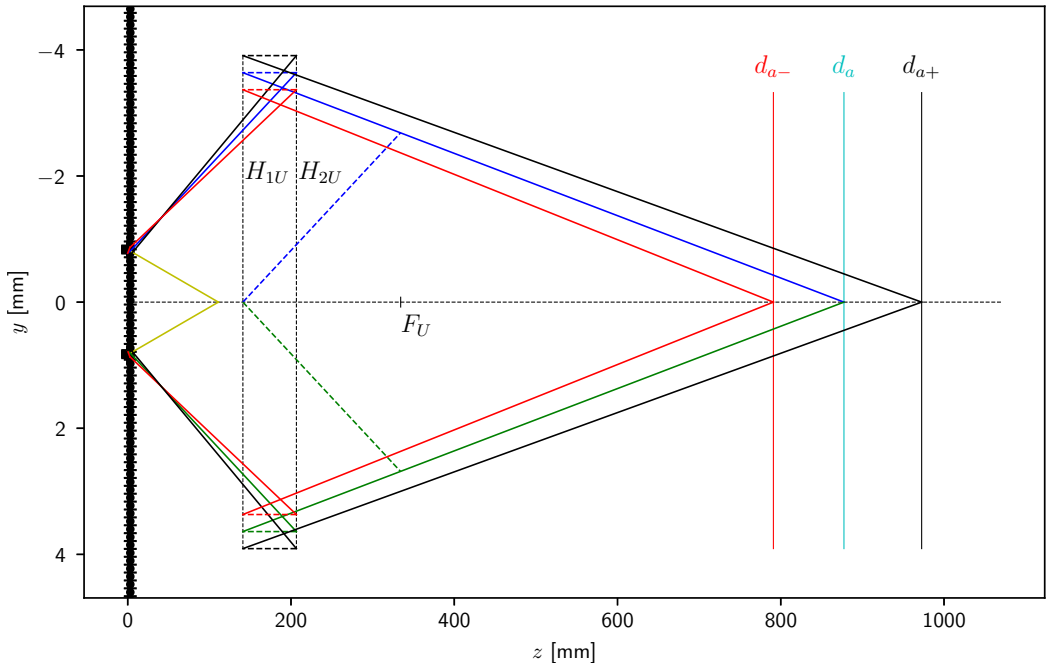}}
	\caption{Refocusing plot for shift parameter $a=1$ with refocusing distance $d_a$ and depth of field limits $d_{a\pm}$ as well as main lens principal planes $H_{1U}$, $H_{2U}$ and focal plane $F_U$. \label{fig:refo}}
\end{figure}

\begin{figure}[H]
	\centering
	\resizebox{0.715\textwidth}{!}{\input{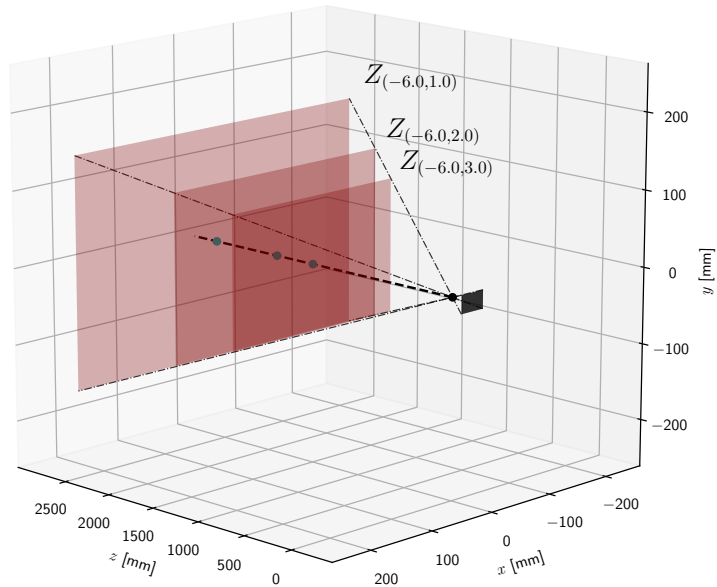}}
	\caption{Triangulation 3-D plot with planes $Z_{(G, \Delta x)}$ and a camera gap $G=-6$\label{fig:3d_tria}}
\end{figure}

Light-field geometry results are provided as text values \if For the command line and graphical user interface, \fi or graphical plots \if by means of \Package{Matplotlib} \fi with two types of views. An exemplary cross-sectional plot is shown in Fig.~\ref{fig:refo}. 
In addition, the results can be displayed in 3-D, as depicted in Fig.~\ref{fig:3d_tria}, with triangulation planes $Z_{(G, \Delta x)}$ for disparities $\Delta x$\if and a default number of depth planes being shown\fi. 
For details on the design trends, scientific notations and deviations due to paraxial approximation, we may refer to our preceding publications~\cite{Hahne:IJCV:18, Hahne:OPEX:16} for further reading.
%

%


\section{Impact}
\label{sec:4}


The presented framework provides answers and generic solutions to a question a researcher raised in a \href{https://www.researchgate.net/post/Does_anyone_know_how_to_estimate_the_depth_in_meters_using_a_Light_Field_Camera_Lytro_Illum}{forum}~\cite{sinan:2017} and received recommendations from peers, indicating demands and potentials for future projects. Research directions that may benefit from this software include experimental photography~\cite{LYTROMAN}, cinematography~\cite{Aggoun:2013} or scientific fields ranging from volumetric fluid particle flow~\cite{Deem_2016} to the many kinds of clinical studies~\cite{prevedel:2014:simultaneous, Li439315, Bedard:17, Palmer:18}.

It may be argued that depth plane localization can be done via extrinsic calibration as traditionally employed in the field of stereo vision. Despite its widespread use, this method cannot be performed prior to the camera manufacturing process, but solely after the fact. The same applies to a calibration of light-field metrics via least-squares fitting~\cite{RemS:16}. Due to the heuristic nature of this approach, zoom lens and optical focus settings as well as temperature fluctuations introduce degrees of freedom to the fitted curve that make calibration sophisticated. After all, it is essential to comprehend and exploit the underlying optical model in the development stage of a plenoptic camera for optimal performance. This part would turn out to be costly with the aforementioned alternatives.


A probably deviant attempt of the presented refocusing distance algorithm has been used in the auspicious Lytro cameras~\cite{LYTROMAN}. However, Lytro's method remains undisclosed up to the present day, which once more underlines the need and practical use of our open-source software.

\section{Conclusions}
\label{sec:5}
Thanks to provided software, the light-field geometry of a plenoptic camera can be accurately predicted with ease of use. This is of crucial interest in the design stage of a prototype where distance range and depth plane density need to be determined and optimized in advance. This tool has been made available for web-servers, a graphical user-interface and the command line tool. It is the first open-source software to do so and may lay the groundwork for future research on plenoptic imaging in the medical field. Future development may lead towards the implementation of an automatic parameter optimization or an extension of a focused plenoptic camera model.



\section*{Acknowledgements}
\label{sec:ack}
We are grateful for the anonymous reviews and Natalie Schnelle for proof-reading this paper. 




\bibliographystyle{elsarticle-num} 
\bibliography{softwarex_article_plenoptisign}

\begin{thebibliography}{10}
\expandafter\ifx\csname url\endcsname\relax
  \def\url#1{\texttt{#1}}\fi
\expandafter\ifx\csname urlprefix\endcsname\relax\def\urlprefix{URL }\fi
\expandafter\ifx\csname href\endcsname\relax
  \def\href#1#2{#2} \def\path#1{#1}\fi

\bibitem{prevedel:2014:simultaneous}
R.~Prevedel, Y.-G. Yoon, M.~Hoffmann, N.~Pak, G.~Wetzstein, S.~Kato,
  T.~Schr{\"o}del, R.~Raskar, M.~Zimmer, E.~S. Boyden, et~al., Simultaneous
  whole-animal 3d imaging of neuronal activity using light-field microscopy,
  Nature Methods 11~(7) (2014) 727.

\bibitem{Li439315}
H.~Li, C.~Guo, D.~Kim-Holzapfel, W.~Li, Y.~Altshuller, B.~Schroeder, W.~Liu,
  Y.~Meng, J.~French, K.-I. Takamaru, M.~Frohman, S.~Jia,
  \href{https://www.biorxiv.org/content/early/2018/10/10/439315}{Fast,
  volumetric live-cell imaging using high-resolution light-field microscopy},
  bioRxiv\href {http://dx.doi.org/10.1101/439315} {\path{doi:10.1101/439315}}.
\newline\urlprefix\url{https://www.biorxiv.org/content/early/2018/10/10/439315}

\bibitem{Bedard:17}
N.~Bedard, T.~Shope, A.~Hoberman, M.~Ann~Haralam, N.~Shaikh, J.~Kovačević,
  N.~Balram, I.~Tošić, Light field otoscope design for 3d in vivo imaging of
  the middle ear, Biomedical Optics Express 8 (2017) 260.
\newblock \href {http://dx.doi.org/10.1364/BOE.8.000260}
  {\path{doi:10.1364/BOE.8.000260}}.

\bibitem{Palmer:18}
D.~W.~Palmer, T.~Coppin, K.~Rana, D.~G.~Dansereau, M.~Suheimat, M.~Maynard,
  D.~A.~Atchison, J.~Roberts, R.~Crawford, A.~Jaiprakash, Glare-free retinal
  imaging using a portable light field fundus camera, Biomedical Optics Express
  9 (2018) 3178.
\newblock \href {http://dx.doi.org/10.1364/BOE.9.003178}
  {\path{doi:10.1364/BOE.9.003178}}.

\bibitem{AW}
E.~H. Adelson, J.~Y. Wang, Single lens stereo with a plenoptic camera, IEEE
  Transactions on Pattern Analysis and Machine Intelligence 14~(2) (1992)
  99--106.

\bibitem{Hahne:IJCV:18}
C.~{Hahne}, A.~{Aggoun}, V.~{Velisavljevic}, S.~{Fiebig}, M.~{Pesch},
  \href{https://doi.org/10.1007/s11263-017-1036-4}{Baseline and triangulation
  geometry in a standard plenoptic camera}, International Journal of Computer
  Vision 126~(1) (2018) 21--35.
\newblock \href {http://dx.doi.org/10.1007/s11263-017-1036-4}
  {\path{doi:10.1007/s11263-017-1036-4}}.
\newline\urlprefix\url{https://doi.org/10.1007/s11263-017-1036-4}

\bibitem{NG}
R.~Ng, Digital light field photography, Ph.D. thesis, Stanford University (July
  2006).

\bibitem{Isaksen:2000:DRL:344779.344929}
A.~Isaksen, L.~McMillan, S.~J. Gortler,
  \href{http://dx.doi.org/10.1145/344779.344929}{Dynamically reparameterized
  light fields}, in: Proceedings of the 27th Annual Conference on Computer
  Graphics and Interactive Techniques, SIGGRAPH '00, ACM Press/Addison-Wesley
  Publishing Co., New York, NY, USA, 2000, pp. 297--306.
\newblock \href {http://dx.doi.org/10.1145/344779.344929}
  {\path{doi:10.1145/344779.344929}}.
\newline\urlprefix\url{http://dx.doi.org/10.1145/344779.344929}

\bibitem{Plenopticam:2019}
{Christopher Hahne}, {PlenoptiCam}, \url{https://github.com/hahnec/plenopticam}
  (April 2019).

\bibitem{Hahne:OPEX:16}
C.~Hahne, A.~Aggoun, V.~Velisavljevic, S.~Fiebig, M.~Pesch,
  \href{http://www.opticsexpress.org/abstract.cfm?URI=oe-24-19-21521}{Refocusing
  distance of a standard plenoptic camera}, Opt. Express 24~(19) (2016)
  21521--21540.
\newblock \href {http://dx.doi.org/10.1364/OE.24.021521}
  {\path{doi:10.1364/OE.24.021521}}.
\newline\urlprefix\url{http://www.opticsexpress.org/abstract.cfm?URI=oe-24-19-21521}

\bibitem{LYTROMAN}
{Lytro Inc.}, {Lytro ILLUM User Manual},
  \url{https://s3.amazonaws.com/lytro-corp-assets/manuals/english/illum_user_manual.pdf},
  2.1 edn. (August 2015).

\bibitem{sinan:2017}
{Sinan Hasirlioglu}, {ResearchGate},
  \url{https://www.researchgate.net/post/Does_anyone_know_how_to_estimate_the_depth_in_meters_using_a_Light_Field_Camera_Lytro_Illum}
  (January 2017).

\bibitem{Aggoun:2013}
A.~Aggoun, E.~Tsekleves, R.~Swash, D.~Zarpalas, A.~Dimou, P.~Daras, P.~Nunes,
  L.~Soares, Immersive {3D} holoscopic video system, MultiMedia, IEEE 20~(1)
  (2013) 28--37.
\newblock \href {http://dx.doi.org/10.1109/MMUL.2012.42}
  {\path{doi:10.1109/MMUL.2012.42}}.

\bibitem{Deem_2016}
E.~A. Deem, Y.~Zhang, L.~N. Cattafesta, T.~W. Fahringer, B.~S. Thurow,
  \href{https://doi.org/10.1088%2F0957-0233%2F27%2F8%2F084003}{On the
  resolution of plenoptic {PIV}}, Measurement Science and Technology 27~(8)
  (2016) 084003.
\newblock \href {http://dx.doi.org/10.1088/0957-0233/27/8/084003}
  {\path{doi:10.1088/0957-0233/27/8/084003}}.
\newline\urlprefix\url{https://doi.org/10.1088%2F0957-0233%2F27%2F8%2F084003}

\bibitem{RemS:16}
R.~{Schima}, H.~{Mollenhauer}, G.~{Grenzd{\"o}rffer}, I.~{Merbach},
  A.~{Lausch}, P.~{Dietrich}, J.~{Bumberger}, Imagine all the plants:
  Evaluation of a light-field camera for on-site crop growth monitoring, Remote
  Sensing 8 (2016) 823.
\newblock \href {http://dx.doi.org/10.3390/rs8100823}
  {\path{doi:10.3390/rs8100823}}.

\end{thebibliography}


\begin{thebibliography}{00}


\end{thebibliography}


\section*{Required Metadata}
\label{}

\section*{Current code version}
\label{}


\begin{table}[!h]
\begin{tabular}{|l|p{6.5cm}|p{6.5cm}|}
\hline
\textbf{Nr.} & \textbf{Code metadata description} & \textbf{Please fill in this column} \\
\hline
C1 & Current code version & 1.1.1 \\
\hline
C2 & Permanent link to code/repository used for this code version & \url{https://github.com/hahnec/plenoptisign} \\
\hline
C3 & Legal Code License & GNU GPL-3.0 \\
\hline
C4 & Code versioning system used & Git \\
\hline
C5 & Software code languages, tools, and services used & Python, JavaScript/AJAX \\
\hline
C6 & Compilation requirements, operating environments \& dependencies & \Package{NumPy}, \Package{Matplotlib}, \Package{tkinter}, \Package{ddt}, \Package{cgi} \\
\hline
C7 & If available Link to developer documentation/manual & \url{https://github.com/hahnec/plenoptisign/blob/master/README.rst} \\
\hline
C8 & Support email for questions & info [{\"a}t] christopherhahne.de \\
\hline
\end{tabular}
\caption{Code metadata (mandatory)}
\label{} 
\end{table}

\newpage
\section*{Current executable software version}
\label{}


\begin{table}[!h]
\begin{tabular}{|l|p{6.5cm}|p{6.5cm}|}
\hline
\textbf{Nr.} & \textbf{(Executable) software metadata description} & \textbf{Please fill in this column} \\
\hline
S1 & Current software version & 1.1.1 \\
\hline

S2 & Permanent link to executables of this version  & \url{https://github.com/hahnec/plenoptisign/releases} \\
\hline
S3 & Legal Software License & List one of the approved licenses \\
\hline
S4 & Computing platforms/Operating Systems & OS X, Microsoft Windows, Unix-like, web-based etc. \\
\hline
S5 & Installation requirements \& dependencies & \Package{NumPy}, \Package{Matplotlib}, \Package{tkinter} \\
\hline
S6 & If available, link to user manual - if formally published include a reference to the publication in the reference list & \url{https://github.com/hahnec/plenoptisign/blob/master/README.rst} \\
\hline
S7 & Support email for questions & info [{\"a}t] christopherhahne.de \\
\hline
\end{tabular}
\caption{Software metadata (optional)}
\label{} 
\end{table}

\end{document}